\documentclass{article}
\usepackage{graphicx}
\usepackage{subcaption}
\usepackage{amsmath,amsthm,amsfonts}
\usepackage{psfrag}
\usepackage{fullpage}
\usepackage{hyperref}
\usepackage{xcolor}
\usepackage{authblk}
\usepackage{placeins}

\newcommand{\ket}[1]{| #1 \rangle}

\newcommand{\braket}[2]{\langle #1 | #2 \rangle}

\newcommand{\comment}[1]{}
\newcommand{\hide}[1]{}

\renewcommand{\phi}{\varphi}

\usepackage{bm}
\usepackage{tabularx}
\newcounter{protocol}
\newenvironment{protocol}[1]{\par\addvspace{\topsep}
  \refstepcounter{protocol}
   \noindent
   \tabularx{\linewidth}{@{} X @{}}
   \hline
   \textbf{Protocol \theprotocol} #1 \\ 
   \hline}{ \\
   \hline
   \endtabularx
   \par\addvspace{\topsep}
 }
\usepackage{enumitem}

\title{Toward multi-purpose quantum communication networks: from theory to protocol implementation}
\author[1,2]{Lucas Hanouz}
\author[1]{Marc Kaplan}
\author[1]{Jean-S\'ebastien Kersaint Tournebize}
\author[1]{Chin-te Liao}
\author[1]{Anne Marin}
\affil[1]{VeriQloud, France}
\affil[2]{Sorbonne Universit\'e, CNRS, LIP6, Paris, France}

\date{}

\begin{document}

\maketitle

\begin{abstract}

Most quantum communication networks around the world are used for a single task: quantum key distribution. In order to initiate the transition to multi-purpose quantum communication networks, we demonstrate the implementation of two different tasks on the same quantum key distribution hardware. Specifically, we focus on quantum oblivious transfer and quantum tokens. Our main contribution is to establish a methodology that greatly simplifies the expertise required to achieve the deployment, assess its performance, and evaluate its feasibility at a large scale.

The implementation that we present is full-stack. It is based on a development framework that allows running user-defined applications both with simulated or real quantum communication backend. The hardware used for the implementation is VeriQloud's Qline. The simulation backend reproduces exactly the inputs and outputs of the real hardware, but also its losses and errors. It can therefore be used to validate the implementation before running it on the real hardware. The sources of the software that we use are fully open, making our research reproducible.

The security of the implementations on real hardware are discussed with respect to security bounds previously known in the literature. We also discuss the engineering choices that we made in order to make the implementations feasible. By establishing a methodology to evaluate the performance and security of quantum communication protocols, we take a significant step towards industrializing and deploying large-scale, multi-purpose quantum communication networks.

\end{abstract}

\section{Introduction}
Quantum communication is a subfield of quantum technologies, which considers the exchange of quantum information between distant agents. Building on the seminal work of Wiesner on quantum money~\cite{Wiesner83}, Bennet and Brassard showed that quantum communication allows two parties to establish a shared secret key with unconditional security~\cite{BB84}. This task, called quantum key distribution (QKD), is regarded as the foundation of quantum cryptography. QKD demonstrates that quantum systems can realize a task that is impossible to achieve using only classical resources. 

Over the years, the research in quantum cryptography has mostly focused on making QKD practical and more secure.
Several protocols have been introduced~\cite{qkd}, record distances have been achieved~\cite{TFQKD23}, and many companies were formed to commercialize the technology. 
QKD devices are now available as off-the-shelf components that can be integrated into classical networks to strengthen their security. The deployment stack using hardware security modules and encryptors is rather well established, at least for telecom grade deployments.

The maturation of the QKD market has been accelerated by the deployment of many QKD testbeds  all over the world. In China, a large-scale QKD network has been deployed progressively since the mid 2010's, even comprising a satellite link. The Micius satellite has even been used to demonstrate trans-continental quantum networks~\cite{micius}. In the EU, the EuroQCI initiative is funding several national projects, which has resulted in the deployment of testbeds in all countries. This has a fostered a strong QKD industry with many start-up companies providing QKD equipments at various degrees of maturity. These testbed deployments play a key role in the maturation of the technology, resulting in improvements of performances and stability of the hardware, and on the other hand, an increase of stakeholders expertise.

In parallel, researchers have investigated a number of different tasks for quantum networks, showing that the range of potential applications goes beyond QKD. Some applications, such as \emph{blind quantum computing} require quantum computers~\cite{vubqc}, while some require simpler pieces of hardware such as entanglement distribution or quantum memories. This induces a hierarchy of quantum networks in terms of hardware available, each stage unlocking new quantum communication tasks~\cite{WEH18}. Furthermore, each new stage expands the range of potential end-users, drawing a clear roadmap from QKD networks to a large scale quantum internet.  This naturally raises the question of expanding single-application quantum networks to multi-purpose ones. 

We initiate the study of this question by considering the implementation of several tasks for quantum networks on the same hardware. Firstly, Implementation leads to benchmarking, a crucial step for the identification of market opportunities for quantum networking technologies. Like for QKD, it requires a large set of skills. Proving security bounds of protocols involves techniques from theoretical quantum cryptography whereas implementing protocol involves the setting of a quantum photonics experiment. Secondly, this approach leads to a unified methodology to implement protocols, from security bounds to practical implementation.

Most quantum information processing tasks cannot yet be implemented with off-the-shelf components. This is the case, for example, of quantum memories that are not integrated yet into quantum communication testbeds. For this reason, most quantum protocols can only be evaluated
 using quantum network simulators~\cite{LBSK22}. 
While this approach can provide feedback on parameters required to demonstrate the security of some communication task, but remains at a low technology readiness level, far from commercialization.

In our work, we focus on applications that can be implemented with the same operations as QKD.
This task only requires one party to generate single qubits chosen uniformly at random from a small set, and another party to immediately measure those states upon receiving them. This is arguably the simplest possible task for quantum communication networks. It only involves two parties, and uses no entanglement or storage. The hardware required for the implementation is mostly standard telecom equipment which exists off-the-shelf.

Surprisingly, other quantum communication protocols that require the same hardware as QKD are much less mature. Protocols such as quantum authentication~\cite{WDKA15}, quantum oblivious transfer~\cite{OT_in_MiniQcrypt} or quantum tokens~\cite{Qtokens} have only been implemented in research labs. Although these implementations can be performed on existing commercial equipment, adapting QKD hardware to run new quantum communication tasks remains particularly challenging.

Moreover, implementing a new quantum communication protocol involves a large set of skills, from experimental quantum photonics to applied cryptography. In order for the executed protocols to be secured, the conditions of the experiment shall satisfy certain constraints expressed in the scientific literature as \emph{security bounds}. Understanding security bounds and verifying that a given execution satisfies those requires a deep interaction between the theory and the experiment. Each protocol has a different security bound, so that executing several protocols using the same experimental setup might be extremely challenging.

To address this issue, we introduce a framework to analyze and evaluate the performances of quantum communication protocols that can run using the same hardware than QKD. This framework rests on a fully programmable quantum communication hardware that allows the implementation of protocols as user-defined applications. To demonstrate the power of our approach, we implement two protocols beyond a simple QKD execution.

The first protocol that we implement is Quantum Oblivious transfer (QOT)~\cite{Kilian88}. In this protocol, two distant parties establish a specific correlation that allow them to evaluate any distributed function securely. This means that the parties don't need to reveal more information about their inputs than what can be deduced from the output of the computation. While oblivious transfer can be computed using only classical resources under standard computational assumption, improving its security with quantum communication is the focus of a long line of work. This started with a proposed protocol with unconditional security that happened to be incorrect~\cite{Bennett92}. Several correction of the protocol were proposed~\cite{DFLSS,Unruh10,BF12}. Recently, a protocol has been proved to be secure in a composable security framework~\cite{OT_in_MiniQcrypt}, assuming only the existence of one-way functions.

The second protocol we implement is for quantum tokens~\cite{Qtokens}. Wiesner introduced Quantum Money, a quantum information processing task achieving a form of security which is impossible to obtain using only classical resources~\cite{Wiesner83}. This had led to a large number of developments (including QKD which is based on Weisner's quantum money scheme). Quantum tokens are, in a nutshell, quantum money without storage. Their security is based on the impossibility to clone quantum states, but they are more restricted in terms of verifiability than general quantum money schemes.

In both cases, we analyze the requirements in term of security, using previously known security bounds. We first simulate the protocols in order to optimize the parameters used in the real implementation. Then, the protocols are executed on the real hardware. This allows to assess the usability of the protocols without changing any of the photonic components, and characterize the current bottlenecks for large-scale deployments.

Our implementation is on the Qline, VeriQloud's open-source hardware~\cite{qline}. For the first step of the process,
we have developed a simulator that reproduces the outputs of the real hardware. Moreover, whenever a user script runs correctly on the simulator, it runs on the real hardware without any change. Strictly speaking, our software thus emulates the Qline hardware. Moreover, we publish all the sources of our emulator and protocols to make our work verifiable and repeatable~\cite{Qline-applications}.

By implementing applications beyond quantum key distribution, we demonstrate that quantum networks can already evolve toward general application networks rather than be limited to quantum key distribution. Our workflow involves the development of new applications on an emulator. This piece of software ensures a smooth deployment onto real-world hardware. Finally, measuring the performances of the implemented protocols improves the characterization of current bottlenecks, a crucial step to explore new use-cases and build a long-term roadmap for quantum communication networks.
Finally, by applying the same methodology to implement several protocols, we take a step toward the transition from single-application quantum networks to multi-purpose ones.

\section{Preliminaries}

\subsection{Quantum information processing and quantum key distribution}
We present some basic notions about quantum information processing and cryptography required to understand the protocols we implement. Readers interested in a more detailed presentation might refer to standard textbooks. We also introduce in the section the notations used throughout our work.

We use bold script $\bm{x}$ for binary strings, that is, $\bm{x} \in \{0,1\}^N$ for some integer length $N$. The $i$-th index of $\bm{x}$ is denoted by $x_i$.
We write $\bm{x} \overset{\$}{\longleftarrow} S$ when the value $\bm{x}$ is chosen uniformly at random from a set $S$. Although in this case, $\bm{x}$ is formally a random variable, we use the same notation as for strings.

All the protocols considered here use BB84 states in a Hilbert state of dimension 2:

\[
\ket 0 = \left ( \begin{array}{c} 1 \\0 \end{array}\right ),
\ket 1= \left(\begin{array}{c} 0 \\1 \end{array}\right),
\ket + =\frac 1 {\sqrt 2} \left(\begin{array}{c} 1 \\1 \end{array}\right),
\ket - =\frac 1 {\sqrt 2} \left(\begin{array}{c} 1 \\-1 \end{array}\right).
\]
The computational basis $\{\ket0, \ket 1\}$ can be mapped to the Hadamard basis $\{\ket +, \ket - \}$ with Hadamard transform $H$, and vice versa.  
For some string $\bm{x} \in \{0,1\}^n$, $\ket{\bm{x}^{\bm{\theta}}} = \bigotimes_{i=0}^{n-1} H^{\theta_i}\ket{x_i}$ is the associated string of BB84 states.

For the protocol considered here, it is also sufficient to consider measurements in the computational basis $\{ \ket 0, \ket 1\}$ or the Hadamard basis $\{ \ket +, \ket -\}$. The outcome of the measurements are labeled 0 and 1 corresponding to the first and second element of the measurement basis, respectively. The probability of obtaining each outcome is the square of the absolute value of the inner product between the state and the vector corresponding to the measurement outcome.

Measuring $\ket 0$ or $\ket 1$ in the computational basis leads to outcome 0 and 1, respectively, with certainty. On the other hand, measuring the same states in the Hadamard basis leads to uniformly random outcomes. Conversely, measuring $\ket +$ or $\ket -$ in the computational basis leads to uniformly random outcomes, while measuring in the Hadamard basis leads to 0 and 1, respectively, with certainty.

Using BB84 sates and the two measurements presented above is sufficient to implement quantum key distribution. The protocol consists in two phases. First, one of the two parties, Alice, sends to the other one, Bob, a sequence of BB84 states chosen uniformly at random.  Upon receiving those, Bob measures in a basis chosen at random between computational and Hadamard. This quantum communication phase is then followed by a post-processing phase that extracts secure keys from the information about the states and measurement results.

Some parts of the post-processing stage of QKD are also used in the other protocols we consider. We give a brief description of those classical sub-protocols, and explain their implementation in more details in the Section~\ref{sec:primitives}.

\textbf{Basis reconciliation or key sifting} consists for Alice and Bob in exchanging information about the bases chosen for states and measurements. In quantum key distribution, they will discard all information corresponding to qubits for which they made different choices. At this stage, the information shared by Alice and Bob is called the \emph{raw key}.

\textbf{Parameter estimation} consists in calculating the average error rate of quantum communication over a small set of data sampled at random. An error happens when the value encoded by Alice and the measurement outcome obtained by Bob are different, although they have chose the same basis. Such error can occur due to any interaction between the quantum communication channel and its environment, including potential eavesdropping.
Intuitively, quantum physics certifies that if an adversary has extracted information from the states, it will induce some errors in the measurement. Parameter estimation is thus used to quantify the maximum information that an adversary can obtain, considering the noise induced in the protocol. 

\textbf{Error Correction} consists in applying an error correction code to correct errors that occured during the transmission of the state, leading Alice and Bob to have the exact same strings.

\textbf{In Privacy amplification}, Alice and Bob use a two-universal hash function to reduce the amount of information that a third party may have obtained about Alice and Bob's private information, while also reducing the size of their keys. The dimension of the resulting string depends on the size of the hash, which is chosen depending on the parameter estimation. Privacy amplification is designed and parametrized so that the adversary can only obtain negligible information about Alice and Bob's shared string after that step, with overwhelming probability.

\subsection{Cryptographic primitives and procedures}
\label{sec:primitives}

Quantum key distribution and the other protocols considered here rely on quantum communication, but also on a number of classical cryptographic primitives. We describe those primitives and the implementation choices we made in our simulations and practical implementations.

\textbf{Pseudo Random Generator (PRG)} are used in the quantum oblivious transfer protocol that we implemented to expand the shared key generated during the execution of the protocol, to an encryption key of arbitrary length. 
We implemented the PRG with SHAKE-256 \cite{Shake256}. This hash function has a variable length output and offers at most $256$ bits of security.

\textbf{Bit commitment}  involves two parties and is decomposed in two phases. In the \emph{commit} phase, Bob commits to some value and sends its commitment to Alice. In the \emph{reveal phase}, Bob opens the commitment and Alice verifies it. In this context, security means that Bob cannot change the commited value after the commit phase, while Alice learns nothing about the committed value before the reveal phase. In quantum protocols, bit commitment is often used to ensure that Alice has measured the quantum states sent by Bob and hasn't stored them until later rounds of the protocols~\cite{DFLSS}. 

We used the following scheme based on SHA-256 hash function~\cite{SHA256}:
%% M: I don't know what are HS and BS. Isn't it confusing???
\begin{itemize}
\item To commit to a bit $b$, Bob chooses a random nonce $r \in \{0,1\}^{\lambda_{HS}}$ and computes the commitment $t = \text{SHA-256}(b || r)$, where $||$ denotes concatenation. He eventually truncates the output to $\lambda_{BS} \leq 256$ bits and sends $t$ to Alice.\\
 We write the procedure to commit to $b$ as $Comm(b) = (t,r) \in \{0,1\}^{\lambda_{BS}} \times \{0,1\}^{\lambda_{HS}}$, and we further write $Pub\circ Comm(x) = t$ to refer to the commitment and public value $t$ and $Priv\circ Comm(x) = r$ to refer to the decommitment and private value $r$. 
\item To reveal the committed bit, Bob sends $(b, r)$ to Alice. Alice verifies the commitment by checking if $t$ matches $\text{SHA-256}(b || r)$.
 We write the verification procedure as:
 
  $\forall t\in\{0,1\}^{\lambda_{BS}}, \forall d = (r,b)\in \{0,1\}^{\lambda_{HS}}\times\{0,1\}, \ Verif(t,d) =
\begin{cases}
1 & \text{if $t = \text{SHA-256}(b || r)$}, \\
0 & \text{else. } 
\end{cases}$
  \item We extend $Comm : \{0,1\} \rightarrow \{0,1\}^{\lambda_{BS}} \times \{0,1\}^{\lambda_{HS}},$ to $n$ values by defining  $Comm(\bm{x}) = (Pub\circ Comm(x_i))_{i=0}^n, (Priv\circ Comm(x_i))_{i=0}^n)$, $\forall \bm{x} \in \{0,1\}^n$.
    \item We define the verification procedure  over $n$ commitments $\bm{Verif}$:
    
      $\forall \bm{t}, \bm{d}\in  {(\{0,1\}^{\lambda_{BS}} \times \{0,1\}^{\lambda_{HS}} \times \{0,1\})}^n, \bm{Verif}(\bm{t},\bm{d}) = \sum_0^{n-1} Verif(t_i,d_i)$  
\end{itemize}

\textbf{Error Correction} of the rawkey is performed using standard classical error-correcting codes. This process reveals some information about the key. As stated by the Shannon's bound \cite{Sha48}, to correct a key of length $n$ transmitted through a binary symmetric channel with crossover probability $p$ (the probability of a bit-flip), one must transmit at least $n h(p)$ bits of information, where $h$ is the binary entropy function $h(x) = -x\log_2(x)-(1-x)\log_2(1-x)$.

We use LDPC codes to perform the error correction. We don't explain the full mathematics behind LDPC codes, but interesting reader can find them in the literature. Operationally, the procedure works as follows:
\begin{itemize}
\item Alice computes the syndrome $\sigma(x)$ of her rawkey $x$ with $\sigma(x) = H x$ where $H$ is the parity check matrix of the code and sends $\sigma(x)$ to Bob.
\item Bob uses a decoder $Decoder$ to compute : $x_{guess} = Decoder(H,p,\tilde{x},\sigma(x))$ where $\tilde{x}$ is the rawkey measured by Bob (after sifting). 
\end{itemize}

We use open-source libraries to implement the error correction. 
The decoder is provided by the \emph{ldpc} Python library \cite{ldpc-decoder} and the parity-check matrices are taken from \emph{LDPC4QKD}~\cite{ldpc4qkd}. This library provides two families of parity-check matrices, whose syndrome length (i.e., the number of leaked bits), are equal to $0.33 n$ and $0.5 n$ respectively, for different block lengths $n$.

The library also proposes a rate adaptation technique in order to further reduce the leakage. We call ``inefficiency'' of the code the ratio: $\frac{|\text{syndrome}|}{n h(p)}$. Due to their probabilistic and iterative decoding algorithm, LDPC codes may exhibit a non-negligible failure probability, referred to as the block error rate (BER). There is a trade-off between low inefficiency and low BER.

\textbf{Privacy Amplification}
Privacy Amplification can be done using a family of $2-$universal hash functions \cite{CW79, WC81}. In our implementation, privacy amplification is achieved via multiplication by a Toepliz matrix \cite{Kra95}.

\section{Quantum communication protocols beyond QKD}

\subsection{Quantum Oblivious Transfer}

Oblivious Transfer (OT), first introduced by Rabin~\cite{Rabin81}, is a fundamental cryptographic primitive used for secure multi-party computation~\cite{Kilian88}. The most commonly used variant, 1-out-of-2 OT (1-2 OT)~\cite{EvenGoldreichLempel85}, involves two parties: a sender Alice with messages of arbitrary length $(m_0, m_1)$, and a receiver Bob with a choice bit $b \in \{0,1\}$. The protocol ensures that Bob receives only $m_b$ while learning nothing about $m_{1-b}$, and Alice remains unaware of Bob's choice $b$.

Oblivious transfer can be used to securely compute distributed functions. In Secure multiparty computing,  the function's inputs are distributed among various parties and security means that no party reveals more information about its input than what can be deduced from the function's output. For two parties, that can be done by using the circuit description of the function and use 1-2 OT to perform distributed multiplications. More elaborate constructions  lead to more efficient MPC implementations based on OT~\cite{IPS08}.

After the introduction of the BB84 QKD protocol, a natural question arose: what other crypto primitives can be implemented with quantum communication better than with classical communication only.  Better may be interpreted in various ways: more efficiently, or more securely. In an attempt to address this question, Brassard and co-authors introduced quantum protocols for bit commitment (BC)~\cite{QBC90} and oblivious transfer~\cite{Bennett92}. Unfortunately, a few years later, these were proven to be insecure~\cite{LC97},  like when using classical communication.

The famous no-go theorem for quantum secure multiparty computing did not close all doors. Oblivious Transfer is known to be a stronger primitive than bit commitment, in the sense that it is possible to use oblivious transfer in order to build any secure multiparty computation primitives~\cite{Kilian88, OT_universality_GMW}, including bit commitment. The opposite, namely building OT from BC, is, however, known to be impossible with classical communication only~\cite{Impagliazzo_Luby,Impagliazzo_Rudich}.

It has first been noticed that the protocol of Bennet, Brassard, Crepeau and Skubiszewska can be fixed using bit commitments~\cite{DFLSS,Unruh10}.
This already indicates that quantum information can perform a task that is not possible with classical communication only.
An explicit security bound for this protocol was later introduced~\cite{BF12}, from which the physical parameters required for security can be derived.

Recently, oblivious transfer was proven to be possible with composable security using only one-way functions~\cite{OT_in_MiniQcrypt}. This shows that secure multiparty computing is possible in a weak cryptographic model called \emph{miniqcrypt}~\footnote{The name refers to Implagliazzo's Minicrypt in which one-way functions exist but we do not have public-key cryptography.~\cite{Impagliazzo}}. However, this proof uses a non-standard construction of bit commitment schemes in order to achieve two properties (equivocality and extractability) required for simulation-based proof of security. In our work, we use a standard, more practical, bit commitment scheme described in Section~\ref{sec:primitives}. 
%We consider the security bound for quantum OT of~\cite{BF12} which has the advantage of being simple compared to the bounds obtained in composable proofs, at the cost of assuming a perfect commitment scheme and dropping composability. 
%While classical OT protocols rely on heavy computational assumptions such as the discrete logarithm problem~\cite{ElGamal85}, quantum OT (QOT) protocols can achieve security from the sole assumption of a secure bit commitment scheme (which can be obtained from a secure One-Way-Function~\cite{Goldreich01}).

We now describe in details the quantum OT protocol that we have implemented.
The protocol is based on Algorithm 5 in~\cite{Y24}, where we replace the ERE-commitment with Protocol 6 in~\cite{ABK23}.

\begin{protocol}{Quantum oblivious transfer}
  \label{protocol:qot}
  \textbf{Input Alice:} two messages $m_0, m_1 \in \{0,1\}^{w}$\\
  \textbf{Input Bob:} $b \in \{0,1\}$\\
  \textbf{Common input:} Alice and Bob agree on security parameters : $\lambda_{OT}, \lambda_{PQS}, \lambda_{BS}, \lambda_{HS}$, and a $PRG : \{0,1\}^{\lambda_{PQS}} \rightarrow \{0,1\}^w$, a commitment scheme $Comm : \{0,1\} \rightarrow \{0,1\}^{\lambda_{BS}} \times \{0,1\}^{\lambda_{HS}}$, associated to a verification procedure $Verif : \{0,1\}^{\lambda_{BS}} \times \{0,1\}^{\lambda_{HS}+1} \rightarrow \{0,1\}$ and $2$-universal hash functions $h_0 : \{0,1\}^{n_0} \times \{0,1\}^{n_0+\lambda_{PQS}} \rightarrow \{0,1\}^{\lambda_{PQS}}$, $h_1 : \{0,1\}^{n_0} \times \{0,1\}^{n_1+\lambda_{PQS}} \rightarrow \{0,1\}^{\lambda_{PQS}}$, where $n_i = |I_i|$ and $I_i$ is defined in Step~\ref{step7}.\\
  \textbf{Output Bob:} $m_b$ or ABORT\\

  \begin{description}
    \item[State distribution]
  \begin{enumerate}[itemsep=0pt, parsep=0pt, topsep=2pt]
  \item Alice chooses $\bm{x}_A \overset{\$}{\longleftarrow} \{0,1\}^{2\lambda_{OT}}$, $\bm{\theta}_A \overset{\$}{\longleftarrow} \{0,1\}^{2\lambda_{OT}}$ and sends $\ket{{\bm{x}_A}^{\bm{\theta}_A}}$ to Bob.
  \item Bob chooses the measurement basis $\bm{\theta}_B \overset{\$}{\longleftarrow} \{0,1\}^{2\lambda_{OT}}$, denotes the results $\bm{x}_B \in\{0,1\}^{2\lambda_{OT}}$.
  \end{enumerate}
\item[Commitment Phase]
  \begin{enumerate}[itemsep=0pt, parsep=0pt, topsep=2pt]
    \setcounter{enumi}{2}
  \item Bob sends $\bm{t} = Pub\circ Comm(\bm{x}_B) \ \| \ Pub \circ Comm(\bm{\theta}_B) $ to Alice.
  \item Alice chooses a random subset $\mathcal{T} \subset [2\lambda_{OT}]$ of size $\lambda_{OT}$ and sends $\mathcal{T}$ to Bob.
  \item Bob sends $\bm{d} = Priv\circ Comm({\bm{x}_B}_{|_{\mathcal{T}}}) \ \| {\bm{x}_B}_{|_{\mathcal{T}}} \ \|  Priv\circ Comm({\bm{\theta}_B}_{|_{\mathcal{T}}}) \ \| \ {\bm{\theta}_B}_{|_{\mathcal{T}}}$ to Alice.
 \item Alice checks that $1 - \bm{Verif}(\bm{t},\bm{d}) / \lambda_{OT} \leq Q_{\text{tol}}$, otherwise, Alice aborts.  
  \end{enumerate}
\item[Basis Reconciliation]
  \begin{enumerate}
     \setcounter{enumi}{5}
  \item Alice sends $\bm{\theta}_A$ to Bob
  \item \label{step7} Bob computes $I_b = \{i \notin \mathcal{T} \ \vert \ \theta_{A_i} = \theta_{B_i}\}$ and $I_{\bar{b}} = \{i \notin \mathcal{T} \ \vert \ \theta_{A_i} \neq \theta_{B_i}\}$ and sends $I_0, I_1$ to Alice
  \item Alice denotes the restriction of $\bm{x}_A$ to $I_i$ : ${\bm{x}_A}_i={\bm{x}_A}_{|_{I_i}}$ for $i\in \{0,1\}$.
  \end{enumerate}
\item[Error Correction]
  \begin{enumerate}
     \setcounter{enumi}{8}
   \item \label{step9} Alice chooses $\bm{s}_0, \bm{s}_1 \overset{\$}{\leftarrow} \{0,1\}^{n0 + \lambda_{PQS}}$ and sends $\sigma_{A0}=Synd({\bm{x}_A}_0)$,  $\sigma_{A1}=Synd({\bm{x}_A}_1)$, $e_0 = PRG(h_0({\bm{x}_A}_0, \bm{s}_0)) \oplus m_0$, $e_1 = PRG(h_1({\bm{x}_A}_1, \bm{s}_1)) \oplus m_1$  to Bob
     %% plus eventually a hash of the keys.
  \item Bob computes $\bm{x}_{guess} = Correct({\bm{x}_B}_{|_{I_b}}, \sigma_{Ab})$ and decrypt ${m_b}_{guess} = PRG(h_b(\bm{x}_{guess},s_b)) \oplus e_b $
   \end{enumerate}
  \end{description}
\end{protocol}

\subsection{Quantum Token}
Around 1970, Wiesner \cite{Wiesner83} proposed the concept of quantum money, exploiting the no-cloning theorem of quantum mechanics to create \emph{unforgeable banknotes}, preventing any duplication or double-spending. However, this early version was a secret-key system, meaning only the bank could verify the money's authenticity, and it required reliable long-term quantum memory to store the qubits, a technology that is way out of reach, even today.

In a nutshell, Wiesner's scheme considers attaching to each banknote a number of photons, each polarized in a random basis. The bank issuing the money keeps a record of the states attached to each banknote. The verification simply consists in measuring each photon in the correct basis and check that the results are consistent with the bank's record. The no-cloning principle ensures that no adversary can create copies of a banknote.

Building on Wiesner's foundation, public-key quantum money was introduced to allow any entity with a quantum computer to verify a token without needing a secret key from the issuer~\cite{Aaronson12} .
The problem of long-term storage of quantum states, however, remained the main blocking point toward the implementation of such protocols. Both public and private-key quantum money scheme require to store quantum states into the banknote until the verification phase begins.

To overcome the limitations of quantum memories, Kent proposed the concept of S-money (\emph{Summonable money})~\cite{Smoney}. In addition to the laws of quantum physics, S-money leverages constraints of special relativity to allow an implementation on existing Quantum Key Distribution infrastructures while retaining security as well as a fair number of interesting aspects of quantum money schemes. 

S-money is a construction that allows a bank to provide a client with digital tokens (banknotes) that can later be redeemed at any one of several predefined presentation points. The protocol ensures instantaneous validation upon presentation, while preventing double-spending (e.g. at multiple locations) and ensuring that the specific presentation point remains unknown to the bank until the transaction is finalized.

In the protocol, the verification agents located at the predefined presentation points do not need long-term quantum storage. This makes the protocol feasible with current technology, in particular on quantum communication testbeds.  When a transaction occurs, quantum information is sent to different locations and measured immediately upon reception. The verification uses only classical communication between the agents.

%The security of S-money relies both on quantum mechanics, specifically the no-cloning theorem, and special relativity. Space-like separation imply timing constraints due to the fact that information signal cannot travel faster than light. This can be used to certify that information cannot be communicated quickly enough to allow token duplication or double-spending.

We implemented the \emph{Quantum Token} scheme of Kent et al. \cite{Qtokens}, itself based on \cite{Smoney}. This protocol involves two distrustful agents: Bob, who generates a token and transmits it to Alice, and Alice, who seeks to redeem the token at one of several predefined locations called \emph{presentation points}. Since then, this protocol has been implemented in several mature environment~\cite{Pan24} and using commercial hardware~\cite{quant24}.

It is assumed that each party is represented by a trusted agent at each presentation point.
In this context, it is possible to enforce the unclonability of quantum tokens
without storing any  quantum information. Instead, it uses the space-like separation of agents to ensure that some of the events are outside of the causal past of some others. Unlike classical protocols, the verification is instantaneous and does not require a round of communication with the bank whenever the tokens are spent.

\begin{protocol}{Quantum token protocol for two presentation points $Q_0$ and $Q_1$ and perfect channel}

  \textbf{Preparation stage:} Alice and Bob agree on parameter $N \in \mathbb{N}$.
  
\textbf{Stage 1}
\begin{enumerate}
\item Bob chooses $\bm{x}_B \overset{\$}{\longleftarrow} \{0,1\}^N$, $\bm{\theta}_B \overset{\$}{\longleftarrow} \{0,1\}^N$ and sends $\ket{{\bm{x}_B}^{\bm{\theta}_B}}$ to Alice.
\item Alice chooses the measurement basis $\bm{\theta}_A \overset{\$}{\longleftarrow} \{0,1\}^N$, denotes the results $\bm{x}_A$.
\item Alice sends $\bm{x}_A$ to $\mathcal{A}_i$ for $i \in \{0,1\}$.
\item Alice chooses $z\overset{\$}{\longleftarrow} \{0,1\}$ and sends Bob the string $\bm{d}$ such that $d=\bm{\theta}_A$ if $z=0$ and $d=\bm{\bar \theta}_A$ if $z=1$.
\item Bob sends $\bm d$, $\bm{x}_B$ and $\bm{\theta}_B$ to $\mathcal B_i$ for $i \in \{0,1\}$. The agents $\mathcal B_i$ for $i \in \{0,1\}$ compute $\bm{d_0} = \bm d$ and $\bm {d_1}=\bm{\bar d}$, respectively, in the causal past of $Q_i$.
\end{enumerate}
\textbf{Stage 2}
\begin{enumerate}
  \setcounter{enumi}{5}
\item Alice chooses the presentation point $p \overset{\$}{\longleftarrow} \{0,1\}$ and sends $c=p \oplus z$ to Bob.
\item Bob sends $c$ to $\mathcal{B}_i$ for $i \in \{0,1\}$.
\item In the causal past of $Q_i$, agent $\mathcal B_i$ computes $\tilde{\bm d_i} = \bm{d_i}$ if $c=0$ and $\tilde{\bm d_i} = \bm{\bar d_i}$ if $c=1$.
\item Alice sends a message to $\mathcal A_p$ to ask to present the token $\bm x_A$ at presentation point $Q_p$.
\item Bob validates the token if $\bm x_A= \bm x_B$ for all $j$ such that Alice's measurement basis is the same as Bob's preparation basis.
\end{enumerate}

\end{protocol}

\begin{protocol}{Quantum token protocol for two presentation points $Q_0$ and $Q_1$ and noisy channel}
\textbf{Preparation stage:} Alice and Bob agree on parameters $N \in \mathbb{N}$ and $\gamma_{det}, \gamma_{err} \in (0,1)$.

\textbf{Stage 1}
\begin{enumerate}
\item Bob chooses $\bm{x}_B \overset{\$}{\longleftarrow} \{0,1\}^N$, $\bm{\theta}_B \overset{\$}{\longleftarrow} \{0,1\}^N$ and sends $\ket{{\bm{x}_B}^{\bm{\theta}_B}}$ to Alice.
\item Alice chooses the measurement basis $\bm{\theta}_A \overset{\$}{\longleftarrow} \{0,1\}^N$, denotes the results $\bm{x}_A \in \{0,1, \emptyset\}^{N}$.
  \item Alice sends the position measured $\Lambda =\{j | {x_A}_j \neq \emptyset\}$ to Bob. Bob aborts if $|\Lambda| < \gamma_{det} N$. 
\item Alice filters $\bm{x}_A$ and $\bm{\theta}_A$ over the received qubits : $\bm{x}_A = ({x_A}_i)_{i \in \Lambda}$ and $\bm{\theta}_A = ({\theta_A}_i)_{i \in \Lambda}$.
\item Alice sends $\bm{x}_A$ to $\mathcal{A}_i$ for $i \in \{0,1\}$.
\item Alice chooses $z\overset{\$}{\longleftarrow} \{0,1\}$ and sends Bob the string $\bm{d}$ such that $d=\bm{\theta}_A$ if $z=0$ and $d=\bm{\bar \theta}_A$ if $z=1$.
\item  Bob filters $\bm{x}_B$ and $\bm{\theta}_B$ over $\Lambda$ : $\bm{x}_B = ({x_B}_i)_{i \in \Lambda}$, $\bm{\theta}_B = ({\theta_B}_i)_{i \in \Lambda}$ and  sends $\bm d$, $\bm{x}_B$ and $\bm{\theta}_B$ to $\mathcal B_i$ for $i \in \{0,1\}$. The agents $\mathcal B_i$ for $i \in \{0,1\}$ compute $\bm{d_0} = \bm d$ and $\bm {d_1}=\bm{\bar d}$, respectively, in the causal past of $Q_i$.
\end{enumerate}
\textbf{Stage 2}
\begin{enumerate}
  \setcounter{enumi}{5}
\item Alice chooses the presentation point $p \overset{\$}{\longleftarrow} \{0,1\}$ and sends $c=p \oplus z$ to Bob.
\item Bob sends $c$ to $\mathcal{B}_i$ for $i \in \{0,1\}$.
\item In the causal past of $Q_i$, agent $\mathcal B_i$ computes $\tilde{\bm d_i} = \bm{d_i}$ if $c=0$ and $\tilde{\bm d_i} = \bm{\bar d_i}$ if $c=1$.
\item Alice sends a message to $\mathcal A_p$ to ask to present the token $\bm x_A$ at presentation point $Q_p$.
\item Bob validates the token if $d({{\bm{x}_A}_|}_{\Delta}, {{\bm{x}_B}_|}_{\Delta}) \le |\Delta|\gamma_{err}$, where $\Delta = \{j \,|\, \tilde{d}_{\bm{p},j} = \theta_{B,j}\}$ is the set of matching bases and $d$ is the Hamming distance. 
\end{enumerate}

\end{protocol}

The transition from 2 to $2^M$ presentation points represents a shift from a binary spatial choice to a multidimensional address framework. While for $M=1$, the protocol uses a single secret bit $z \in \{0,1\}$, the generalized version leverages an $M$-bit secret string $\mathbf{z} \in \{0,1\}^M$ to navigate an expanded configuration space. To maintain logical alignment between the short address string and the larger sequence of $NM$ quantum pulses, the protocol introduces a repetition function $\text{rep}(\cdot)$ that extends each bit $z_k$ over a block of $N$ pulses, facilitating a rigorous commitment $\mathbf{d} = \theta_A \oplus \text{rep}(\mathbf{z})$. This architectural scaling ensures that the reconstruction of preparation bases $\tilde{d}_p$ remains strictly localized to the specific agent $\mathcal{B}_p$ corresponding to Alice's choice $p \in \{0,1\}^M$, thereby preserving unforgeability and user privacy across $2^M$ potentially spacelike separated regions.

\begin{protocol}{Quantum token protocol for $2^M$ presentation points and noisy channel}

\textbf{Preparation stage:} Alice and Bob agree on the number of address bits $M \in \mathbb{N}$, the number of pulses per block $N \in \mathbb{N}$ (total pulses $NM$) and thresholds for detection $\gamma_{det}$ and error $\gamma_{err} \in (0,1)$.

\textbf{Stage 1}
\begin{enumerate}
\item Bob chooses $\bm{x}_B \overset{\$}{\longleftarrow}\{0,1\}^{NM}$ and preparation bases $\bm{\theta}_B \overset{\$}{\longleftarrow} \{0,1\}^{NM}$, then sends the quantum states $\ket{{\bm{x}_B}^{\bm{\theta}_B}}$ to Alice.
\item Alice chooses random measurement bases $\bm{\theta}_A  \overset{\$}{\longleftarrow} \{0,1\}^{NM}$, measures the pulses, and records the results $\bm{x}_A \in \{0,1, \emptyset\}^{NM}$.
\item Alice  sends the position measured $\Lambda =\{j | {x_A}_j \neq \emptyset\}$ to Bob. Bob aborts if there exists $l\in [0,M-1]$ s.t $|\Lambda_{|j\in\{lN, \ldots,(l+1)N-1\}} |< \gamma_{det} N$. 
\item Alice filters $\bm{x}_A$ and $\bm{\theta}_A$ over the received qubits : $\bm{x}_A = ({x_A}_i)_{i \in \Lambda}$ and $\bm{\theta}_A = ({\theta_A}_i)_{i \in \Lambda}$
\item Alice distributes the results $\bm{x}_A$ to all her local agents $\mathcal{A}_i$ for $i \in \{0,1\}^M$ via secure channels.
\item Alice chooses a secret string $\bm{z} \overset{\$}{\longleftarrow} \{0,1\}^M$. She sends Bob the commitment string $\bm{d}$ such that $\bm{d} = \bm{\theta}_A \oplus \text{rep}(\bm{z})$, where $\text{rep}(\bm{z})$ extends each bit $z_l$ of $\bm z$ over the $|\Lambda_{|j\in\{lN, \ldots,(l+1)N-1\}}|$ received pulses.

 \item Bob filters $\bm{x}_B$ and $\bm{\theta}_B$ over $\Lambda$ : $\bm{x}_B = ({x_B}_i)_{i \in \Lambda}$, $\bm{\theta}_B = ({\theta_B}_i)_{i \in \Lambda}$ and sends $\bm d$, $\bm{x}_B$ and $\bm{\theta}_B$ to $\mathcal B_i$ for $i \in \{0,1\}^M$. The agents $\mathcal B_i$ computes $\bm{d_i} = \bm{d_i} \oplus \text{rep}(\bm{i})$.
\end{enumerate}

\textbf{Stage 2}
\begin{enumerate}
  \setcounter{enumi}{6}
\item Alice selects a presentation point $\bm{p} \in \{0,1\}^M$ and sends the choice string $\bm{c} = \bm{p} \oplus \bm{z}$ to Bob.
\item Bob broadcasts $\bm{c}$ to all agents $\mathcal{B}_i$.
\item Each agent $\mathcal{B}_i$ computes the reconstructed bases $\tilde{\bm{d}}_i = \bm{d}_i \oplus \text{rep}(\bm{c})$. %Note that for the correct agent $i=\bm{p}$, $\tilde{\bm{d}}_{\bm{p}} = \bm{\theta}_A$.
\item Alice instructs $\mathcal{A}_{\bm{p}}$ to present the token $\bm{x}_A$ at point $Q_{\bm{p}}$.
\item Agent $\mathcal{B}_{\bm{p}}$ validates the token if $d({{\bm{x}_A}_|}_{\Delta}, {{\bm{x}_B}_|}_{\Delta}) \le |\Delta|\gamma_{err}$, where $\Delta = \{j \,|\, \tilde{d}_{\bm{p},j} = \theta_{B,j}\}$ is the set of matching bases and $d$ is the Hamming distance. 
\end{enumerate}

\end{protocol}

\section{The software stack}

The applications are deployed on VeriQloud's quantum communication hardware Qline~\cite{qline}. In order to support application development, we designed a layered software stack that facilitates the development of quantum communication protocols.
This structure was designed to enable a development workflow in two steps. Firstly, the applications are developed on a simulator that replicates the outputs of the real hardware. Secondly, the same applications are executed on the real quantum communication hardware to validate them and measure their performances in real conditions.

Our approach is not to be confused with a quantum network stack. While we recognize that such a stack would be helpful and certainly necessary for future large scale deployments, it goes beyond the scope of our work. Our goal in this section is only to present the software we have used in order to accelerate the deployment of user-defined protocols on existing quantum communication hardware.

\subsection{Stack description}
The stack consists in three layers deployed by both communicating parties in order to run user-defined applications. The complete software stack is open and can be downloaded to reproduce the results presented here.
We describe the stack in more details. We don't fully describe the technical processes used, but interested readers can directly read through the source code~\cite{Qline-applications}.

\textbf{At the Hardware Layer}, the users may choose between two different backends. The other software of the stack can be either executed on the Qline quantum communication devices or on a hardware simulator.
The simulator, called \textbf{hwsim}, reproduces exactly Qline's inputs and outputs.
For this reasons, we may think of the software as a hardware emulator but it also contains advanced simulation features that allow to tune hardware parameters such as losses and detection rate. In the protocols presented here, the inputs of the two parties are independent random strings, and the outputs are correlated random strings corresponding to a prepare-and-measure protocol using BB84 states.

\textbf{The Global Counter} layer's role it to maintain the synchronization between the communicating parties. The global counter refers to the indices given to emitted and measured pulses. This layer is oblivious of the fact that the pulses were physically produced or logically simulated.
It ensures that both parties identify the bits corresponding to inputs and outputs consistently.
We refer to this layer as~\textbf{gc}.

\textbf{The Application} layer includes protocols built on top of quantum communication. In this paper, we consider specifically two applications: oblivious transfer and quantum tokens. In practice, any protocol in which Alice sends random BB84 states and Bob measures them in a randomly chosen basis can be implemented, including of course QKD.

\begin{figure}[!ht]
    \centering % \hspace*{0.1cm}
    \includegraphics[scale=0.5]{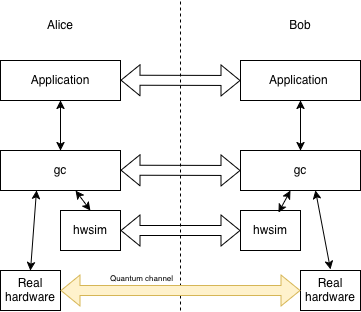}
    % \caption{\centering Performances.}
    \label{fig:softstack}
        \caption{The software stack -- the application interacts with \textbf{gc} but remains oblivious of the backend. It can run both on simulated or real quantum hardware.}
\end{figure}

\subsection{Workflow}

The shared correlated strings are generated by the hardware layer at each party. They correspond to the information provided by Alice when she generates a sequence of BB84 states, and the information obtained by Bob when measuring them.
For the protocols implemented here, these strings contain the basis chosen for encoding and the description of quantum states for Alice, and the measurement basis and results for Bob.  Moreover, the correlation between the strings depends on the qubit error rate, the fiber losses and the detection rate. 

The software \textbf{gc}  runs on both parties. It writes the derived strings into local FIFO buffers, making sure that both parties remain synchronized. 
This design guarantees that application-layer software can continuously read correlated data on both sides in a consistent way. If, in some application, Alice refers to the $i$-th state, Bob can unambiguously identify the corresponding measurement and result.

The software \textbf{hwsim} creates random strings classically, reproducing the correlation obtained when those strings are established using quantum communication. The physical parameters (error rate, fiber losses, etc...) as well as a random seed are stored in a configuration file distributed to both parties. The seed is then expanded to reproduce the results of quantum communication. Notice that this solution offers no security and may only be used to validate the execution of the protocols before their deployment on the real hardware.

In order to provide a complete framework for application development, we also expose a control API that allows applications to actively start or stop the key data stream established by the physical layer.
Conceptually, this API acts as a data-stream switch, controlled via a classical communication channel.
By decoupling data generation from application consumption, this design enables fine-grained control over resource usage and supports dynamic application behavior without compromising synchronization guaranties.

\section{Security bounds and simulations}

\subsection{Quantum Key Distribution}

We first introduce the security bound we use for QKD protocols~\cite{TL17}. This bound gives a standard measure of performances for the hardware used for our implementations. We also use the well-established case of QKD to discuss the methodology for securely implementing a protocol using an appropriate security bound from the literature. Our objective for the other protocols is to reproduce this methodology using some security bounds found in the literature.

In QKD, the security is measured by the distance between the real state $\rho_{KE}$ produced by the protocol and the ideal state $\frac1{2^l}I \otimes \rho_E$. The state obtained in the protocol may result from any physically valid operation performed by an eavesdropper trying to learn information about the key. The state $\rho_{KE}$ thus represents the information that can be extracted about the key. The security bound then indicates how to choose the parameters of the protocol to certify that this information remains small or, in the case of QKD, that the resulting state is close to a separable state between a uniformly random key and the information that the eavesdropper gets.

We consider the following security bound:

\begin{equation}\label{eq:QKD_security_bound}
  \Delta(\rho_{KE} ,\frac1{2^l} I \otimes \rho_E)
    \leq
    \frac12 \times
    2^{
      -\frac12[
        n(1-h(Q_{\text{tol}}+\delta))- r - q - l
      ]
    }
    + 2e^{
      -\frac{nk^2\delta^2}{(n+k)(k+1)},
    }
\end{equation}
where:
\begin{itemize}
\item  $\Delta$ is the trace distance,
\item $n$ is the length of the rawkey and $k$ is the length of the sample extracted from the sifted key for parameter estimation, so that $n+k$ is the length of the sifted key,
\item $Q_{\text{tol}}$ is the maximum qubit error rate tolerated by the protocol,
\item $\delta$ accounts for statistical fluctuations,
\item $r$ is the length of the hash function used to check the correctness of the key,
\item q is the length of the syndrome used to correct the rawkey, and
\item l is the final length of the secret key after performing privacy amplification.
\end{itemize}

By setting $\epsilon_{\text{sec}} =   4e^{-\frac{nk^2\delta^2}{(n+k)(k+1)}}$, $\epsilon_{\text{cor}} = 2^{-r}$ and defining the final length l as a function:
\begin{equation}\label{eq:final_length_qkd}
  l(\epsilon_{\text{sec}}, \epsilon_{\text{cor}}, n, k, Q_{\text{tol}}, q)  =
  n \bigg[
  1-h\Big(
  Q_{\text{tol}}  + \sqrt{\frac{n+k}{nk}\frac{k+1}{k}\ln{\frac{4}{\epsilon_{\text{sec}}}}}
  \,\Big)
  \bigg]
  - q - \log_2{\frac{1}{\epsilon^2_{\text{sec}}\epsilon_{\text{cor}}}} 
\end{equation}
we have from \ref{eq:QKD_security_bound} that :
\begin{equation}
  \Delta(\rho_{KE} ,\frac1{2^l} I \otimes \rho_E) \leq \epsilon_{\text{sec}} 
\end{equation}

In practice, the choice of the parameters in Equation~\ref{eq:final_length_qkd} determines whether the implementation is secure. We give some indications of how the parameters in the security bounds are chosen.

The security parameter $\epsilon_{\text{sec}}$ is chosen reasonably small to make sure that the adversary cannot extract information from the protocol. For example, a security parameter of $10^{-14}$ means that no eavesdropper is able to distinguish the actual key from a random bit string of the same length with higher probability than $\frac12 + 10^{-14}$ (where $\frac12$ is the probability of a random guess).

The parameter $Q_{\text{tol}}$ is an upper bound on the admissible error rate: the protocol will remain secure for any error rate below $Q_{\text{tol}}$. The choice of this parameter depends on the hardware. Ideally, this parameter can be adjusted during the execution of the protocol.

The length of the sample extracted for a correct parameter estimation $k$ is $O(\sqrt n)$~\cite{TL17}. The best factor in $O$ depends on the length, the details of the protocol, and other parameters. It is optimized by each QKD vendors depending on its estimation of the security.
%\textcolor{red}{Suggestion (Anne): to remove $70\sqrt n$ ?, this factor depends on the length, the protocol, etc and could always be optimized. Eg when we are looking for minimum N to extract few bits of secure key like in OT / token etc, k=n is better.}

The length of the syndrome $q$ depends on the error correcting code. A good target is $q=1.2 h(Q_{\text{tol}})$. In our implementation, the factor varies from 1.25 when using Cascade to values between 1.3 to 1.7 (depending on the qber) when using LDPC codes.
%\textcolor{red}{Remark (Anne): 1.2 is for 'good codes', we are trying to reach this value. I think that for Cascade we were around 1.25 and with the LDPC used here in OT, we managed to have 1.3 to 1.7 (depending on the qber), and this is probably possible to do better.}

In QKD protocols, the main goal is to maximize the secret key rate, whether it is defined as the number of secret key bits generated per unit of time or per raw key bit. The bound in Equation~\ref{eq:QKD_security_bound} gets tighter as $n$ grows. Engineering constraints, however, give upper bounds on the block size $n$ that can be processed, so maximizing the secret key rate effectively corresponds to maximizing $n$ under post-processing limitations, such as those imposed by error correction and privacy amplification.

\begin{figure}[!ht]
    \centering % \hspace*{0.1cm}
    \includegraphics[scale=0.4]{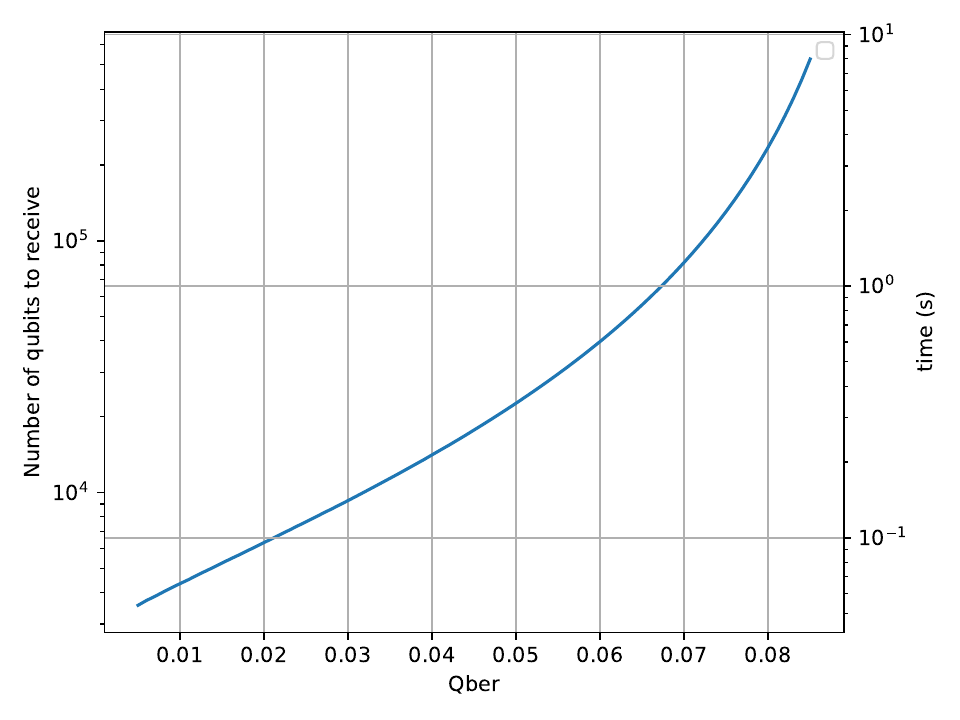}
    \caption{Minimum number of qubits to be received by Bob and corresponding time to extract one bit of secure key, using a good code (inefficiency $1.2$), $\epsilon_{\text{sec}}=10^{-10}$, $\epsilon_{\text{cor}}=10^{-10}$. The hardware parameters are those of Qline.}
    \label{fig:minNqkd}
\end{figure}

In contrast, for other protocols such as quantum oblivious transfer or quantum tokens, increasing the block size once the target security is achieved provides no additional benefit. Thus, the main goal in these cases is to minimize the block size $n$ required to reach the desired security level.  Figure~\ref{fig:minNqkd} shows the minimum number of qubits needed to be received in a QKD protocol to get a positive secret key rate $l/n$ satisfying Equation~\ref{eq:QKD_security_bound}.

%The bound in Equation~\ref{eq:QKD_security_bound} gets tighter as $n$ grows. Engineering constraints, however, give upper bounds on the value $n$ that can be processed. The processing time of error correction, for example, depends on the value $n$. In practice, we fix a value $n$ so that error correction takes a reasonable time.

%After those parameters are fixed, we get the final length of the key $l$ as a function of the length of the rawkey $n$. In Figure~\ref{fig:minNqkd}, we use it  to show, for different values of the qbit error rate, the number of qubits that needs to be received by Bob in order to get a positive keyrate.

\subsection{Quantum oblivious transfer (Q-OT)}

Similar to the QKD protocol,  the security of the Q-OT protocol against a malicious Bob is defined as the distance between a real state and an ideal state. Here the real state is the joint state of Alice's and Bob's systems $\rho_{K_{\bar{b}}K_bE}$, consisting of Alice's private key (or message) $K_{\bar{b}}$, which should remain unknown to Bob, Alice's key $K_b$ to which Bob is allowed access and Bob's environment $E$. The ideal state $\frac1{2^l} I\otimes \rho_{K_b}\otimes \rho_E$ consists in a random $l$-bit key  $\frac1{2^l} I$ independent of the rest of the system  $\rho_{K_b}\otimes \rho_E$.
We used the following security bound~\cite{Y24}~\footnote{In the original bound, other parameters such as potential additional leaks are taken into account. We set them to 0 in our context}:

\begin{equation}\label{eq:OT_security_bound}
    \Delta(\rho_{K_{\bar{b}}K_bE} ,\frac1{2^l} I \otimes \rho_{K_b}\otimes \rho_E) \leq
    \frac12 \times
    2^{
        -\frac12[
            (\frac12-\xi-h(Q_{\text{tol}} + \delta))\frac{\lambda_{OT}}{2}
            - q -l
        ]
    }
    + \sqrt6 e^{
        -\frac{\delta^2}{100} \lambda_{OT}
    }
    + 2 e^{
        -\frac{\xi^2}{2}\lambda_{OT}
    }
\end{equation}
where:
\begin{itemize}
  %  \item $\epsilon$ is the security measure. The maximum probability for a malicious Bob to distinguish Alice's behavior from an ideal simulator without breaking the commitment scheme is $\frac12 + \epsilon$ (see \cite{Y24} for more details).
  \item $\Delta$ is the trace distance
    \item $\xi$ and $\delta$ are margins on the statistical fluctuation emerging from the potential deviations between statistics of samples and of the whole populations. The bound holds for any strictly positive values of $\xi$ and $\delta$ as long as $h(Q_{\text{tol}}+\delta)$ is well defined. $\xi$ and $\delta$ are thus to be chosen accurately to minimize the final value of the security paramter $\epsilon$.
    \item $Q_{\text{tol}}$ is the tolerable qubit error rate.
    \item 2$\lambda_{OT}$ is the number of \emph{received} photons (as claimed by Bob).
    \item $l$ is the output length of the 2-universal hash function used in the privacy amplification step of the post-processing. In our implementation, we fixed $l=256$.
    \item $q$ is the length of the syndrome of the error correcting code used in the error correction step of the post-processing.
\end{itemize}
We further set $\epsilon_{\text{sec1}} =  2\sqrt6 e^{-\frac{\delta^2}{100} \lambda_{OT}}$ and $\epsilon_{\text{sec2}} = 2 e^{-\frac{\xi^2}{2}\lambda_{OT}}$ and define the final length l as a function:
\begin{equation}\label{eq:final_length_ot}
  l(\epsilon_{\text{sec1}}, \epsilon_{\text{sec2}}, \lambda_{OT}, Q_{\text{tol}}, q)  =
  \frac{\lambda_{OT}}{2}
  \Bigg[
  \frac12 - \sqrt{\frac2\lambda_{OT}\ln{\frac4\epsilon_{\text{sec2}}}}
  - h\Big( Q_{\text{tol}} + \sqrt{\frac{100}{\lambda_{OT}}\ln{\frac{2\sqrt6}{\epsilon_{\text{sec1}}}}} \,\Big)
  \Bigg]
   - q - \frac1{(\epsilon_{\text{sec1}}+\epsilon_{\text{sec12}})^2}
\end{equation}

Then, from \ref{eq:OT_security_bound} and \ref{eq:final_length_ot}, we have:
\begin{equation}
  \Delta(\rho_{K_{\bar{b}}K_bE} ,\frac1{2^l} I \otimes \rho_{K_b}\otimes \rho_E) \leq \epsilon_{\text{sec1}} + \epsilon_{\text{sec2}}
\end{equation}

Our target $\lambda_{PQS}$ security for Protocol~\ref{protocol:qot} is $256$ bits, which determines the length of the seed for the PRG in Step~\ref{step9} of Protocol~\ref{protocol:qot}.
Thus, after fixing $\epsilon_{\text{sec1}}$ and $\epsilon_{\text{sec2}}$, the measure of performance for securely running the OT protocol consists in finding the smallest value $\lambda_{OT}$ such that $l(\epsilon_{\text{sec1}}, \epsilon_{\text{sec2}}, \lambda_{OT}, Q_{\text{tol}}, q)\geq\lambda_{PQS}$.

\begin{figure}[!ht]
    \centering % \hspace*{0.1cm}
    \includegraphics[scale=0.4]{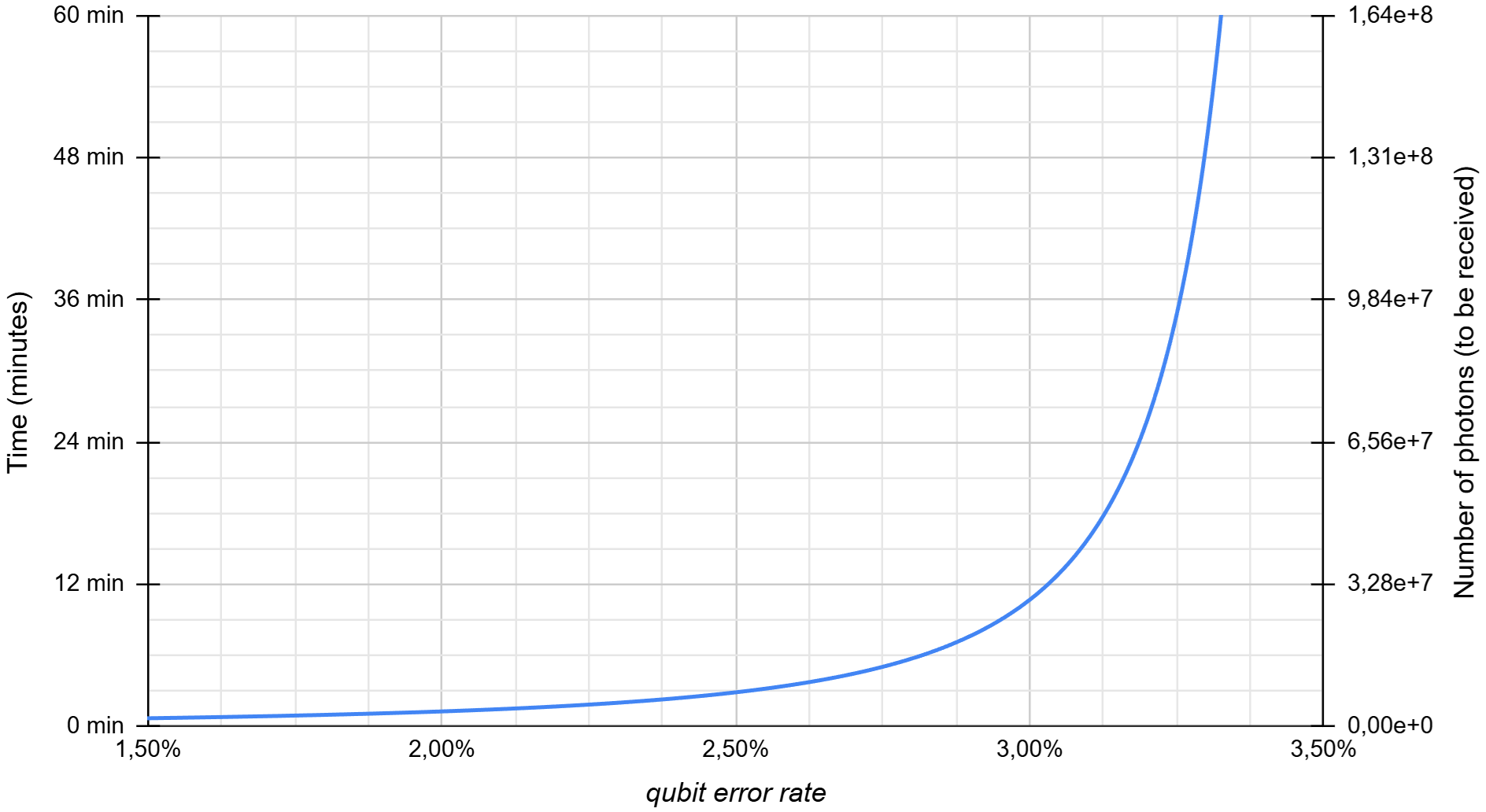}
    \caption{Estimation of the number of qubits to be received, and corresponding time required on Qline, to perform one OT as a function of the qber. The inefficiency of the error correction is set to $1.25$.}
    \label{fig:perf_OT}
\end{figure}

By using a numerical approach to minimize $\epsilon$ over the choices of $\xi$ and $\delta$, we obtained an upper bound on the number of qubits that need to be received in order for the protocol to terminate securely. Figure~\ref{fig:perf_OT} shows, as a function of the qubit error rate, the number of photons that need to be received by Bob for a secure execution of the protocol (right axis), as well as the corresponding expected time (left axis).

%Figure~\ref{fig:perf_OT} represents values obtained by estimating the size of the syndrome to $1.25\times h(Q_{\text{tol}})$, where the binary entropy of the qubit error rate $h(Q_{\text{tol}})$ is the minimum size of the syndrome of a perfect error correction code. %($\times n$ missing). 

\subsection{Quantum tokens}

Practical quantum tokens~\cite{Qtokens} consider a weak coherent source with mean photon number $\mu$. In particular, a dishonest party can take advantage of multiphoton events resulting from the use of weak coherent states, as well as deviation in the prepared states from ideal BB84 states.

The corresponding security bound of the practical quantum token protocol is given by
\begin{equation}\label{eq:QT_security_bound}
  \epsilon_{\text{unf}} \leq
  e^{\frac{P_{\text{noqubit}}N}{3} {\big(\frac{\nu_{\text{unf}}}{P_{\text{noqubit}}} -1\big)}^2}
 +
 e^{-Nf(\gamma_{\text{err}}, \beta_{PS}, \beta_{PB}, \phi, \nu_{\text{unf}}, \gamma_{\text{det}})}
\end{equation}
where:
\begin{align}\label{eq:QT_security_bound_f}
  f(\gamma_{\text{err}}, \beta_{PS}, \beta_{PB}, \phi, \nu_{\text{unf}}, \gamma_{\text{det}})  =
  & (\gamma_{\text{det}} - \nu_{\text{unf}}) \Big[\frac{\lambda(\phi, \beta_{PB})}{2}{\Big(1 - \frac{\delta}{\lambda(\phi, \beta_{PB})} \Big)}^2 - \ln{(1 + 2\beta_{PS})} \Big] \notag\\
  & - (1 - (\gamma_{\text{det}} - \nu_{\text{unf}}))\ln{(1 + h(\beta_{PS}, \beta_{PB}, \phi) )}, 
\end{align}
\begin{equation}\label{eq:QT_security_bound_h}
  h(\beta_{PS}, \beta_{PB}, \phi) = 2\beta_{PS}\sqrt{\frac12 + 2\beta^2_{PB} + (\frac12 - 2\beta^2_{PB})\sin(2\phi)}
\end{equation}
and
\begin{equation}\label{eq:QT_security_bound_delta}
  \delta = \frac{\gamma_{\text{det}}\gamma_{\text{err}}}{\gamma_{\text{det}} - \nu_{\text{unf}}}
\end{equation}
under the following constraints:
\begin{equation}\label{eq:QT_security_bound_cons1}
  0 \leq \gamma_{\text{err}} \leq  \lambda(\phi,\beta_{PB})
\end{equation}
\begin{equation}\label{eq:QT_security_bound_cons2}
  0 \leq P_{\text{noqubit}} \leq  \nu_{\text{unf}} \leq \text{min}\big\{ 2P_{\text{noqubit}}, \gamma_{\text{det}}\Big( 1 - \frac{\gamma_{\text{err}}}{\lambda(\phi, \beta_{PB})} \Big) \big\}
\end{equation}
\begin{equation}\label{eq:QT_security_bound_cons3}
  0 \leq \beta_{PS} \leq \frac12 \Big( e^{\frac{\lambda(\phi, \beta_{PB})}{2}{\big(1 - \frac{\delta}{\lambda(\phi, \beta_{PB})} \big)}^2} - 1 \Big)
\end{equation}

where:
\begin{itemize}
\item $\epsilon_{unf}$ is the probability of successfully forging an unauthorized quantum token,
\item $P_{noqubit}$ is the probability of a multiphoton event. For a weak coherent state of mean photon number $\mu$, we have $P_{\text{noqubit}} = 1 - (1 + \mu) e^{-\mu}$,
\item $N$ is the number of qubits sent,
\item $\gamma_{\text{err}}$ is the chosen threshold for the maximum tolerable qubit error rate,
\item $\beta_{PB}$ is the biais over the basis preparation. $\beta_{PB} = \left| \frac{|{\theta_B}_i \in \bm{\theta_B}\, |\, {\theta_B}_i = 0|}{|\Lambda|}  - \frac12 \right|$, where $\bm{\theta_B}$ is restricted to the set of reported measured pulses $\Lambda$,
\item $\beta_{PS}$ is the biais over the state preparation. $\beta_{PB} = \max\{ \left| \frac{|{x_B}_i \in \bm{x_B}\, |\, {x_B}_i = 0, \, {\theta_B}_i = 0|}{|\Lambda|}  - \frac12 \right|, \left| \frac{|{x_B}_i \in \bm{x_B}\, |\, {x_B}_i = 0, \, {\theta_B}_i = 1|}{|\Lambda|}  - \frac12 \right|\}$, where $\bm{x_B}$ is restricted to the set of reported measured pulses $\Lambda$,
\item $\phi$ is the maximum deviation in state preparation from the ideal BB84 states. For all $\theta, \theta' \in \{0,1\}$, $| \braket{x^\theta_0}{x^{\theta'}_1} | \leq O(\phi) = \frac{1}{\sqrt2} (\cos\phi + \sin\phi) $  where $O(\phi)$ is the maximum overlap between  $\ket{x^\theta_0}$ and $ \ket{x^{\theta'}_1}$,
 \item $\lambda(\phi, \beta_{PB}) = \frac12 \Big( 1 - \sqrt{1 - (1 - \frac{\cos\phi + \sin\phi}{\sqrt2})^2}(1-4\beta_{PB}^2) \Big)$,
\item $\gamma_{\text{det}}$ is the chosen threshold for the minimum detection rate.

\end{itemize}

Depending on the parameters, the number of required photons to be sent for an execution of the Qtoken protocol varies. As a consequence, so does the time required to perform the protocol for a fixed repetition rate. 

\begin{figure}[htbp]
     \centering
     \begin{subfigure}[b]{0.47\textwidth}
         \centering
         \includegraphics[width=\textwidth]{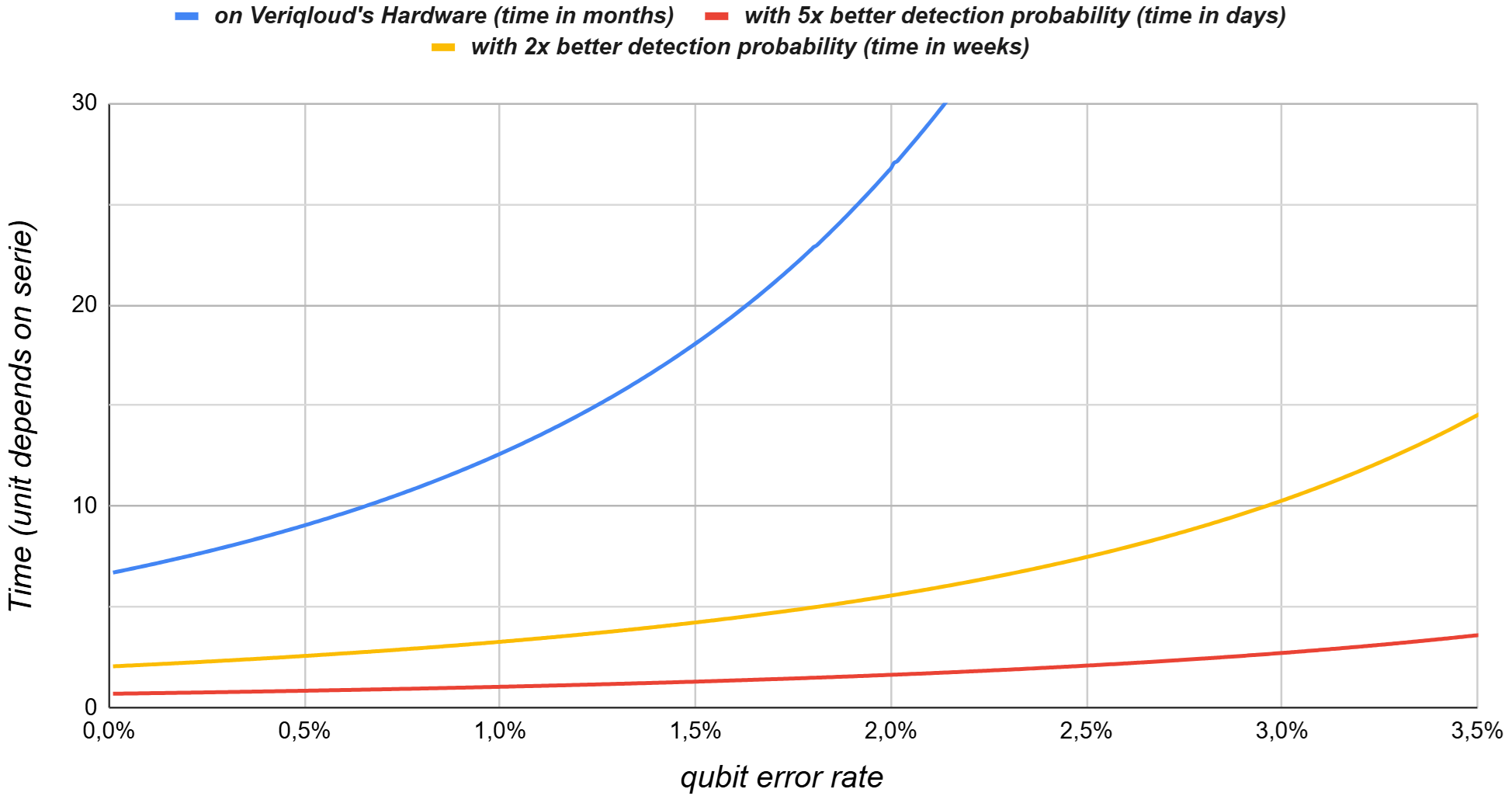}
%         \caption{\centering depending on the qubit error rate}
         \label{fig:Qtokens_perf_qber}
     \end{subfigure}
     \hfill 
     \begin{subfigure}[b]{0.47\textwidth}
         \centering
         \includegraphics[width=\textwidth]{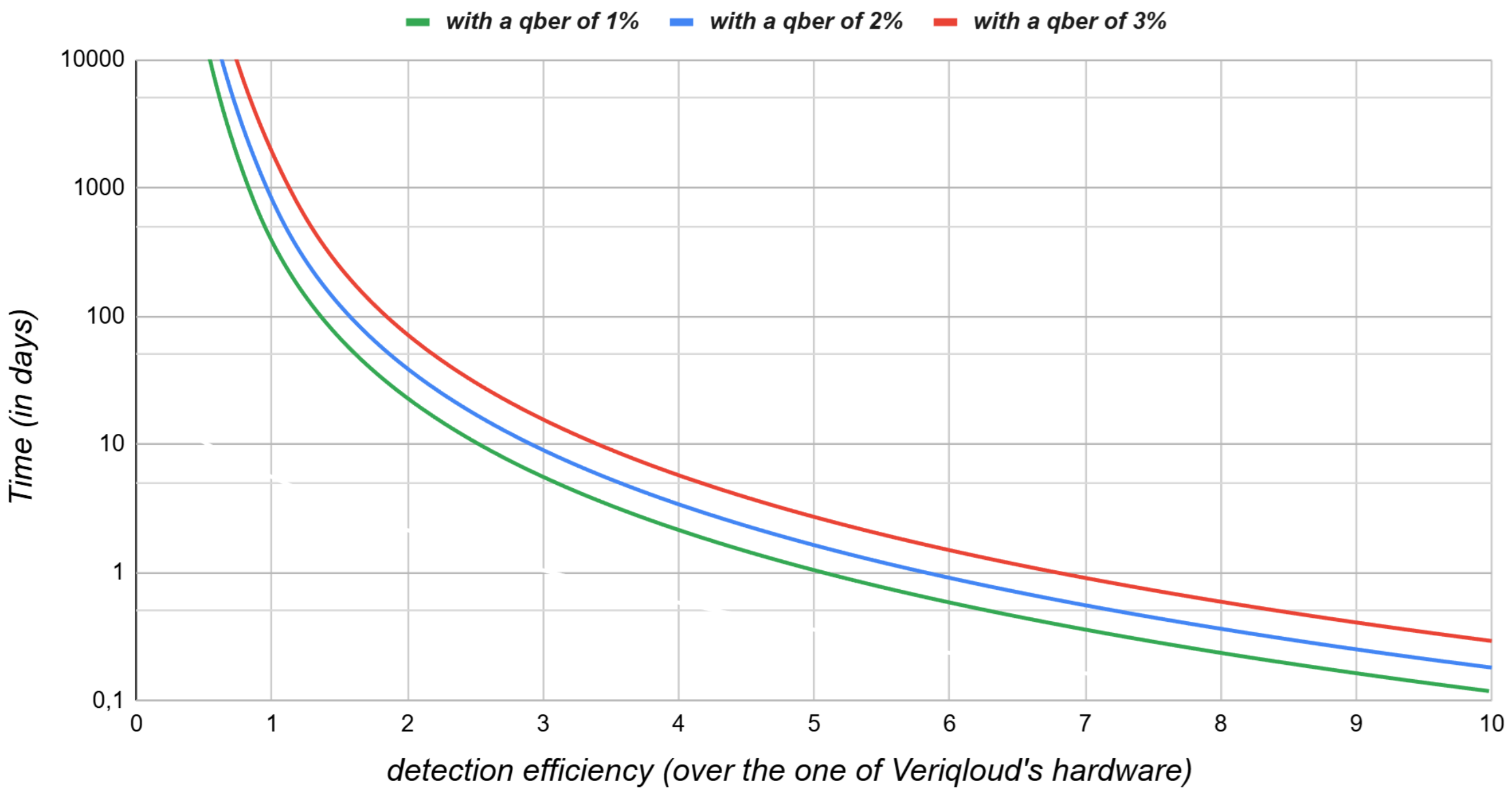}
%         \caption{\centering depending on the detection efficiency}
         \label{fig:Qtokens_perf_pdet}
     \end{subfigure}
     \caption{Time required to generate 1 quantum token depending (a) on the qubit error rate or (b) on the detection efficiency}
     \label{fig:Qtokens_perfs}
\end{figure}

We used the numerical simulation to optimize the number of photons required, over the mathematical variable $\nu_{unf}$ and the range of performance trade-offs achievable on Qline.
Results suggest that the two parameters with the largest effects of the security are, on the one had, the threshold for the receiver's detection probability of an incoming photon (with a large impact) and on the other hand, the threshold $\gamma_{err}$ for the Qubit Error Rate (with a lower impact).
Figure~\ref{fig:Qtokens_perfs} shows the evolution of the required time to perform the Qtoken protocol depending on the probability for Bob to detect an incoming photon (left), and depending on the qubit error rate (right).

% In blue, optimized parameters for VeriQloud's Hardware (with a detection probability of about $0.01$). In red an yellow, optimized parameters for a hardware with respectively a 10 times and 25 times better detection probability. As the values differ in several orders of magnitude, we chose to display each curve in a different unit of time to improve readability.

%simulation à très bonne detection efficiency?

\section{Experimental results}

\subsection{Description of the hardware}

Our protocols are implemented VeriQloud's Qline. This hardware implements quantum key distribution, but as we described earlier, it can also run user-defined applications. The hardware is open-source and 
its internal description is published on repositories accessible from Qline's website~\cite{qline}. We briefly describe the characteristics and performances of the hardware. Qline consists in two boxes, one for Alice and one for Bob. Alice's repetition rate is 80 MHz, with a qber contained between 2 and 7\%. The total loss budget is 25dB.

The photonic hardware included into Qline are, on Alice's side a laser, an amplitude modulator and a phase modulator. The qubits are encoded in the phase difference between pulses in two time bins called \emph{early} and \emph{late}, the two pulses corresponding to a single photon in a superposition of early and late. On Bob's side, we use a phase modulator, an interferometer and avalanche photodiode photon detector to decode the qubit. Using a single APD allows savings on the hardware costs, but the single-photon interference used to read the qubit value induces losses that are inherent to the encoding.

Qline's QKD keyrate has been measured in field~\cite{QlineDT} and in the lab. The measurement results correspond to the keyrate deduced from the security bound. Depending on the attenuation, it varies from 10kbps to 10bps. The performance graph is also included in Qline's documentation.

%For a given set of physical parameters, Qline's keyrate can be deduced from the previous parameters and the security bound that we use. This bound expresses the distance between the random distribution obtained between Alice, Bob and a potential eavesdropper and an ideal distribution, in which Alice and Bob have the same random distribution but an eavesdropper has no information about it. This distance can be made arbitrarily small by using the right amount of privacy amplification, which in turn requires to extend the length of the raw key.

The post-processing is performed on a standard computer with processor Intel i7-14700T, 16 Gb of RAM, a 500Go SSD. This hardware is sufficient to execute the whole post-processing of QKD, but limits operations on large data structures such as the block size of the error correcting code. The computer also runs the software gc mentioned earlier. 

The two Qline devices also exchange classical information at each level of the stack. The classical network is deployed using White-Rabbit switches, which also distribute the clocks between the devices using the White-Rabbit protocol. We haven't identified any bottleneck resulting from network latency. 

\subsection{Quantum OT}

%\subsection{Quantum OT}

The quantum oblivious transfer described in Protocol~\ref{protocol:qot} is executed on Qline.
At the time of the experiment, the average qber was $2\%$. We set the tolerable qber accordingly to $Q_{\text{tol}}=0.025$.
The parity-check matrix we used, described in the LDPC4QKD library~\cite{ldpc4qkd}, is of size $1\,572\,864\times 524\,288$. We applied rate adaptation techniques which reduces the code inefficiency to 1.40 for $Q_{\text{tol}}=0.025$. Using the largest available block allows to apply the rate adaptation while keeping the decoding failure probability low (below 0.06).

We set the security parameters to $\epsilon_{\text{sec1}}=\epsilon_{\text{sec2}}=2^{-23}$, yielding an overall security of $\epsilon_{\text{sec}} = 2^{-22}$. This value is below the typical security level of $10^{-10}$ but allows to reduce the minimum resource requirements by a factor of approximately 1.33.
%% M: unclear what "minimym resource requirements" means

\begin{figure}[!ht]
     \centering

      \begin{minipage}[b]{\textwidth}
     \centering
     \includegraphics[width=0.5\textwidth]{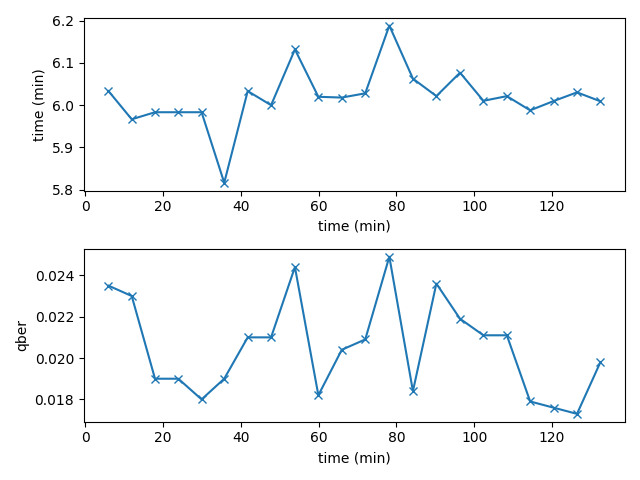}
     \caption{OT running on VQ hardware: measured time to obtain one OT on Qline (top subplot) and evolution of the qber over time (bottom subplot). The number of qubits to be received is set to $N\approx 1.26 \times 10^7$.}
     \label{fig:ot_xp}
     \end{minipage}

     \begin{minipage}[b]{0.45\textwidth}
     \centering
     \includegraphics[width=0.99\textwidth]{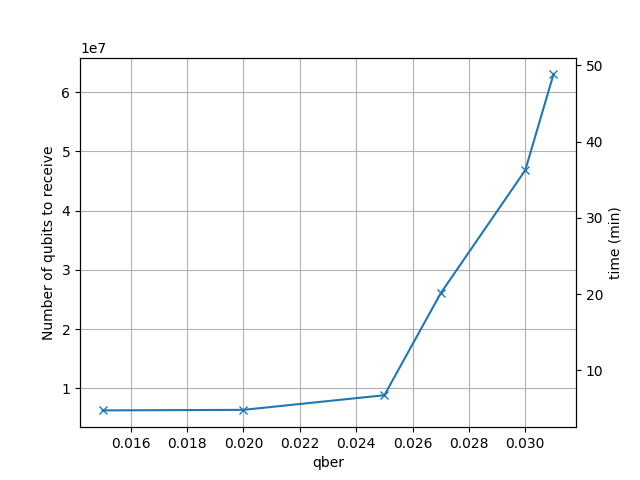}
     \caption{Simulated OT running on Qline: Number of qubits that need to be received (left axis) and corresponding expected time (right axis) to obtain one OT on Qline.}
     \label{fig:ot_simu}
     \end{minipage}
     \hfill
      \begin{minipage}[b]{0.45\textwidth}     
     \centering
     \includegraphics[width=0.99\textwidth]{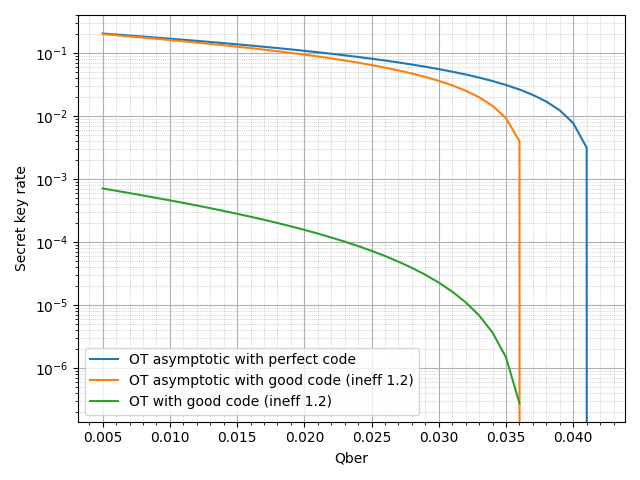}
     \caption{OT ``secret key" rate $l/a$ with $l=256$ and $a=\lambda_{OT}$, the length of non-open commitments.}
     \label{fig:ot_sk_rate}
     \end{minipage}

     \begin{minipage}[b]{\textwidth}
     \centering
     \includegraphics[width=0.5\textwidth]{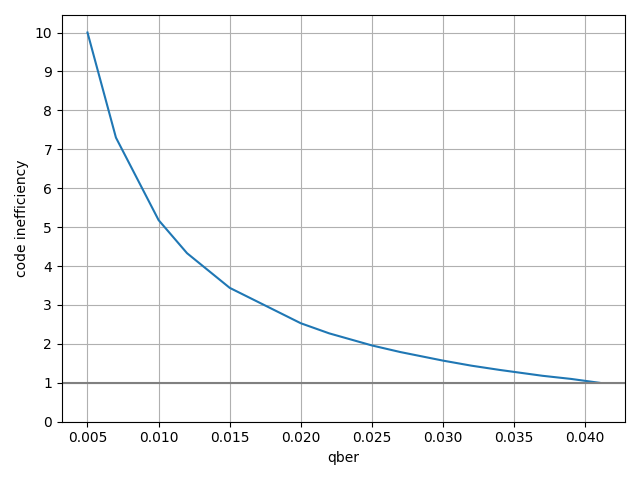}
     \caption{Maximum tolerable code inefficiency to extract 256 bits of secure key with $\epsilon_{\text{sec}}=2^{-34}$.}
     \label{fig:max_ineff}
     \end{minipage}
     
\end{figure}

For $Q_{\text{tol}}=0.025$, using the security bound in Equation~\ref{eq:OT_security_bound} and security parameters $\epsilon_{\text{sec1}}=\epsilon_{\text{sec2}}=2^{-23}$, the shortest of the two potential raw keys ${\bm{x}_A}_0, {\bm{x}_A}_1$ in Protocol~\ref{protocol:qot}  should have length at least $n_0=2\,420\,736$ to extract at least 256 bits of secure key. By setting the number of qubits to be received before post-processing to $N = 2\lambda_{\text{OT}}=12\,595\,200$, we have $|{\bm{x}_A}_i| \geq n_0$ with probability at least 99.999\%
The results of the running experiment are shown in Figure~\ref{fig:ot_xp}.
%further improvement
The average achieved \textit{OT-rate} is 1/6 OT per minute. 

We do not use any parallelism in our implementation. We evaluate that parallelizing the quantum transmission and the post processing could improve the performances by a factor of almost two. This can be seen in Table~\ref{tab:xp_times}, which shows the time to complete each task.

Figure~\ref{fig:ot_simu} shows the execution time of the OT protocol on our simulator as a function of the qber with the same security parameters as in the real experiment and the same code (with two possible rate adaptations, before and after $qber = 0.025$). We were not able to simulate the protocol for qbers above 3.1\%. The limitation is due to memory limitations during the commitment phase. During this phase, Bob commits to $2\lambda_{OT}=N \geq 60\,000\,000$ bits.
 The minimum number of qubits to be received also differs from that depicted in Figure~\ref{fig:perf_OT} by a factor of two. The difference is due both to the variable syndrome size of the implemented LDPC codes, whose inefficiency ranges between $1.31$ and $1.67$ depending on the qber (compared with $1.25$ in Figure~\ref{fig:perf_OT}), and to the reduction in the security parameter, which satisfies $\epsilon_{\text{sec}} = 2^{-22}$ in Figure~\ref{fig:ot_simu} versus $\epsilon_{\text{sec}}< 10^{-10}$ in Figure~\ref{fig:perf_OT}.  %MK: what does that mean?
Further optimization over the code (with the rate-adaptation technique or with different code families) would produce better results. 

Figure~\ref{fig:ot_sk_rate} shows that, in theory, the OT protocol could run securely up to a qber of 3.6\% using a better code with inefficiency 1.2, while the asymptotic limit using a perfect code reaches a qber of 4.1\%.    
In Figure~\ref{fig:max_ineff}, we plot the maximum tolerable inefficiency to run the OT protocol with a security of $\epsilon=2^{-34}$. This limit is theoretical in the sense that the corresponding post-processing block size N is of the order of $10^{11}$, which would make the protocol impractical.

%Security parameters:
%$\lambda_{OT}(\epsilon)$
%Commitment:
%$\lambda_{BS}$ -> 136 bits (taille du hash).
%$\lambda_{HS}$ -> 64 bits (taille des r de decommitment)

\begin{table}
  \centering
  \begin{tabular}{|c|c|c|}
    \hline
    Task & Time (s) & Fraction\\
    \hline
    q-receive & 204 & 56.7\% \\
    \hline
    commitment & 89 & 24.7\%\\
    \hline
    decoding & 51 & 14.2\%\\
    \hline
    PA & 16 & 4.4\%\\
    \hline
    total & 360 & 100\%\\
    \hline
%qber\_measured :0.021,\\
    %left\_error:0\\
  \end{tabular}
  \caption{time repartition by task for a typical run of OT on Qline}
  \label{tab:xp_times}
\end{table}

\FloatBarrier

\subsection{Quantum  Tokens}

As shown on Figure~\ref{fig:Qtokens_perfs}, Qline's performances are not good enough to allow an execution of the Qtokens protocol in feasible time.
The main bottleneck is the detection probability, which is too low in our implementation.
We still provide a proof-of-principle demonstration in which we deliberately run the Qtokens protocol with small numbers of received photons ($10^6$ and $10^5$). This leads to insecure tokens, but generated in practical time.

\begin{figure}[!ht]
     \centering
     \includegraphics[width=0.6\textwidth]{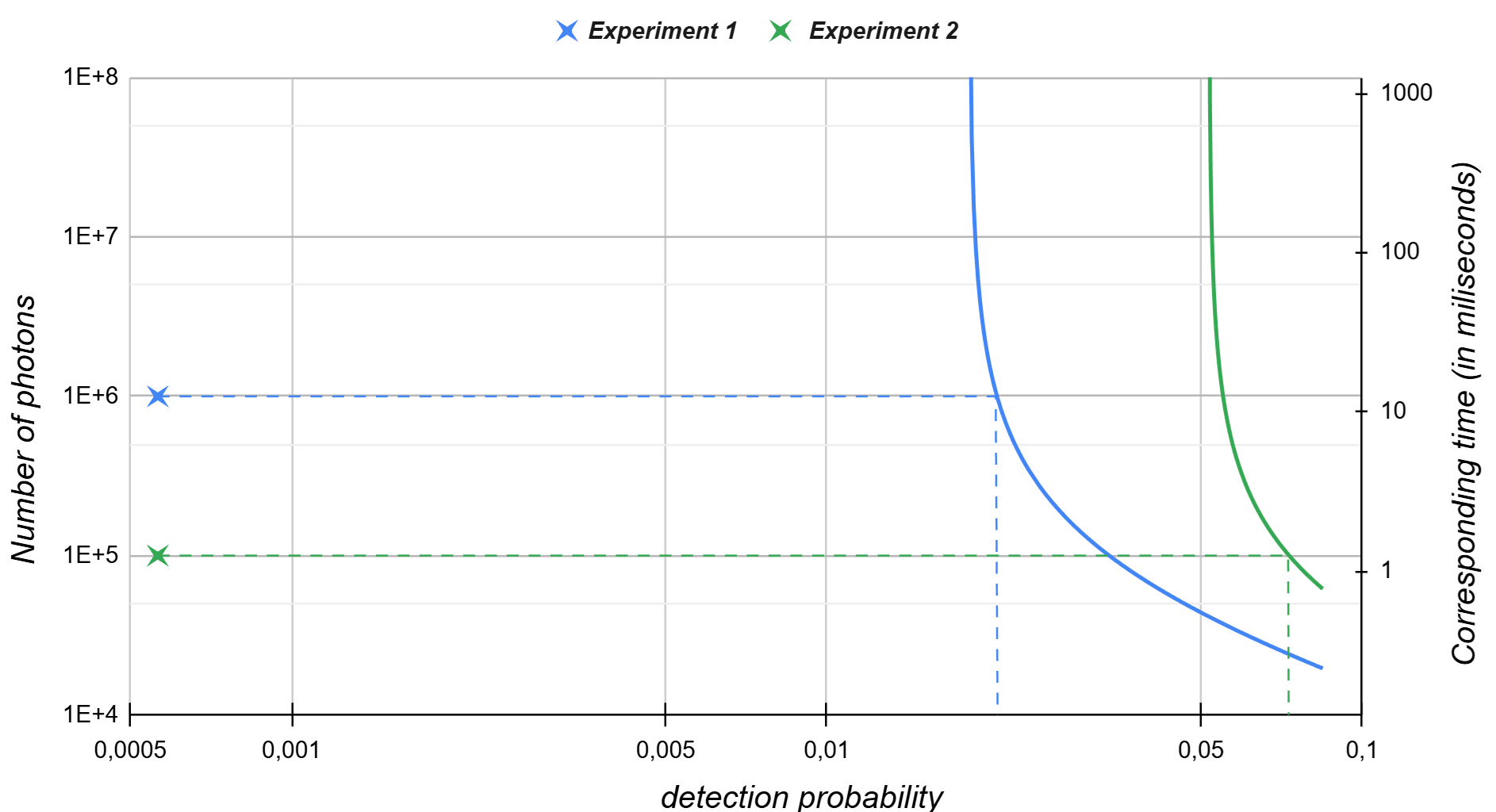}
     \caption{Required number of photons to be received (left axis) and corresponding time (right axis) to obtain one token with $\epsilon_{unf} = 10^{-10}$, depending on the probability that a sent photon is detected, for two different experiments.}
     \label{fig:QT_perf_experiments}
\end{figure}

Among 10 runs of the protocol on Qline, we selected $2$, one with $10^5$ and the other with $10^6$ received photons. OnFigure~\ref{fig:QT_perf_experiments}, they are represented by the crosses on the left of the graph, with a detection probability around $5,6\times10^{-4}$.
For both runs, the observed qubit error rate was around $5.6\%$
%%% Marc: why is the qber so high compared to OT???

For both experiments, Figure~\ref{fig:QT_perf_experiments} shows the number of photons that need to be received to obtain a security (unforgeability $\epsilon_{unf}$) lower than $10^{-10}$, depending on the probability that a photon sent by Alice is successfully detected by Bob.
Since the mean photon number of the weak coherent states used for both experiments was set to $0.1$, this probability cannot exceed $\frac{1}{10}$.
The difference between the two curves is explained by hardware fluctuations between the two runs, and more importantly by different biases $\beta_{PB}$ and $\beta_{PS}$ measured in the different runs.

The dashed lines on  Figure~\ref{fig:QT_perf_experiments} identify the detection probability required for the tokens of each run to be secure. Results show that detection probabilities of $2.08\times10^{-2}$ and $7.29\times10^{-2}$, respectively, would have been required for the each experiments. This  corresponds to detection efficiencies about $37$ and $130$ times higher than the ones obtained on Qline.

%, and to respectively about $21\%$ and $73\%$ of the maximal theoretical efficiency with such a qubit source
We point out, however, that the required detection efficiencies are not particularly out of reach for existing technology.
A setup with a deterministic single photon source like a quantum dot, and an efficient detector such as a superconducting nanowire single-photon detector (SNSPD) could realistically achieve detection probabilities three order of magnitude higher than our setup. Moreover, while the previous experiment in \cite{Qtokens} was also not able to demonstrate a secure token, the detection efficiency of the setup it used was high enough to consider secure tokens in practical time.

%In conclusion, these results validate the operational workflow of the protocol, and indicate that future iterations leveraging a higher-performance hardware could achieve the generation of secure, valid tokens within a reasonable time frame.

\section{Conclusion}

We have implemented two quantum communication protocols beyond QKD on the same quantum communication hardware. The first one, quantum oblivious transfer, is the main cryptographic primitive for secure multiparty computing. The second one, quantum tokens, does not need quantum memories but still guarantees that the tokens cannot be cloned.

Our work demonstrates that standard QKD hardware are not limited to key establishment but can be used for a larger range of protocols. Increasing the number of applications of quantum networks may increase the range of end-users. The fact that it can be done with existing equipments shows that QKD networks can be tuned to run user-defined applications.

This approach also paves the way to a future quantum internet, a network that connects remote quantum objects.
A quantum internet is naturally multi-purpose, using the hardware for realizing different tasks. Our work clearly emphasizes the difficulty in determining a multi-purpose network stack, even in the simple case of prepare and measure protocols. The reason is that there is no uniform framework to securely implement quantum communication. Each security bound takes into consideration different parameters. The security of a protocol implementation is very dependent on the physical parameters of the communication channel.

For QKD, the translation from the theoretical analysis of the security bound into practical physical operations has already been realized. Arguably, each QKD vendor has to operate this translation to ensure that the protocol they implement is secure. In our work, we have expanded this approach to other protocols. Operating, however, in the regime of secure QKD does not translate into the security of other prepare-and-measure protocol. A compiler that translates protocols that are secure in an ideal world into protocols that are secure when considering errors and losses would be an important tool for the development of quantum communication networks.

The security analysis of the protocols that we implemented derive from careful analysis of the security bound. 
Just like QKD, the protocols we implemented require a classical post-processing, which includes in particular error correction and privacy amplification. 
The parameters used in this phase may affect the security in a different way than it does for QKD. There is no unified way to take into account the qber or the losses of the protocol, but rather each protocol requires it own security analysis.

In terms of performances, the  protocols we considered remain impractical even though the keyrate of QKD on Qline is comparable to standard industrial hardware. We were able to general only a few OTs per hour, while with our current hardware it would take several years to obtain a single quantum token.
For each protocol, our implementation help to identify the bottlenecks limiting the rate of resource generation.

Now that the bottlenecks are identified, we can start to think about how to improve those rates. Once again, the solution is specific to each protocol, and there is no unique approach to this question. An easy way would be to improve the performances of the hardware. Here again, the precise physical performance that should be improved depends on the protocol, but arguably, increasing the qubit generation rate would decrease the time of resource generation.

We believe that more theory work can lead to much better security bounds and more efficient implementations. This requires defining new protocols and analyzing their security again. In the case of quantum tokens, one gain that seems reachable is to add decoy states. The current security bounds is obtained to limit the effect of information leakage resulting from photon-number splitting attacks. The purpose of decoy states is exactly to prevent such attacks, and it could be helpful to secure quantum tokens as well.

Security might be easier to achieve as the hardware improves. In the meantime, it is important to work on the theory and identify the best techniques to secure quantum communication protocols in the real world. These techniques seem to apply to several protocols, but a unified framework remains to be defined.

\section{Acknowledgements}
This work was partly funded by the European Union's Horizon Europe research and innovation programme under grant agreement No. 101102140 – QIA Phase~1. We thank Georg Harder, Hop Dinh, Lyes Rekhou and Faten Ben Chabaane for their help in running the hardware. We thank Ruben Cohen for his help with the use of the software stack. We thank Alvaro Yang\"uez for his help in understanding the existing security bounds for OT.

\bibliographystyle{plain}
\bibliography{references}

\end{document}